\documentclass[prd,preprint,a4paper,superscriptaddress,nofootinbib,11pt]{revtex4}
\usepackage{graphicx}
\usepackage{amssymb}
\usepackage{graphicx}
\usepackage{amsmath}
\usepackage{eurosym}
\usepackage{color,graphicx}
\usepackage{dsfont}
\usepackage{amsmath}
\usepackage{amssymb}
\usepackage{amsthm}
\usepackage{stmaryrd}
\usepackage{mathrsfs}
\usepackage{textcomp}
\usepackage{amsfonts}
\usepackage{txfonts}

\def\fs{\footnotesize}
\def\bea{\begin{eqnarray}}
\def\eea{\end{eqnarray}}
\def\be{\begin{equation}}
\def\ee{\end{equation}}
\newtheorem{theorem}{Theorem}
\newtheorem{definition}[theorem]{Definition}
\newtheorem{example}[theorem]{Example}
\newtheorem{proposition}[theorem]{Proposition}
\newtheorem{remark}[theorem]{Remark}

\def\U{\mathcal{U}}
\def\H{\mathcal{H}}
\def\A{\mathcal{A}}
\def\F{\mathcal{F}}
\def\W{\mathcal{W}}

\begin{document}

\title{Bicrossproduct construction versus Weyl-Heisenberg algebra}

\author{A. Borowiec\footnote{borow@ift.uni.wroc.pl}}

\affiliation
{\fs
Institute for Theoretical Physics, University of Wroclaw, pl. Maxa Borna 9,
50-204 Wroclaw, Poland\\}

\author{A. Pacho{\l }\footnote{pachol@raunvis.hi.is}}
\affiliation{{\fs
Science Institute, University of
Iceland, Dunhaga 3, 107 Reykjavik, Iceland}}

\begin{abstract}
We are focused on detailed analysis of the Weyl-Heisenberg algebra in the framework of bicrossproduct construction.
We argue that however it is not possible to introduce full bialgebra structure in this case, it is possible
to introduce non-counital bialgebra counterpart of this construction. Some remarks concerning bicrossproduct basis
for $\kappa-$Poincar\'{e} Hopf algebra are also presented.\end{abstract}
\maketitle

\section{Introduction}
Bicrossproduct construction, originally introduced in \cite{MR_bicross1} (see
also \cite{MR_bicross2}, \cite{M_book} for more details), allows us to construct a new
bialgebra from two given ones. Its applicability to Weyl-Heisenberg algebra is a subject of our study here. In fact, algebraic sector of Weyl-Heisenberg algebra relies on crossed-product construction \cite{Klimyk}--\cite{smash6}
while the coalgebraic one will be main issue of our investigation here. One can easily show that full bialgebra
structure cannot be determined in this case.  However appropriate weakening of some assumptions   automatically  allows on
bicrossproduct type construction.

We start this note with reviewing the notions of Weyl-Heisenberg algebra and indicating its basic properties.
Then we recall definitions of crossed product algebras, comodule coalgebras, their crossed coproduct and bicrossproduct
construction. We follow with some
examples of bicrossproduct construction for the classical inhomogeneous
orthogonal transformations as well as for  the $\kappa-$deformed case.
The coaction map which provides $\kappa-$Poincar\'{e}
quantum (Hopf) algebra \cite{LNRT} was firstly proposed in \cite{MR}. In fact, the system of generators
used in the original construction \cite{MR} which preserves Lorentzian sector algebraically undeformed is called "bicrossproduct
basis". It became the most popular and commonly used by many authors in various applications, particularly in doubly special relativity formalism (see e.g. \cite{DSR1}-\cite{DSR3}) or quantum field theory on noncommutative $\kappa-$Minkowski spacetime (cf. \cite{kappaQFT1}-\cite{kappaQFT4}). However bicrossproduct construction itself is a basis independent. Therefore we also  demonstrate that the so-called classical
basis (cf. \cite{BP2}) leaving entire Poincar\'{e} sector algebraically undeformed is consistent with the bicrossproduct construction and can be used instead as well.

\section{Preliminaries and notation}
Let us start with reminding that Weyl-Heisenberg algebra\footnote{In this note an algebra means unital, associative algebra over a commutative ring which is assumed to be a field of complex numbers $\mathds{C}$ or its h-adic extensions $\mathds{C}[[h]]$ in the case of deformation.}
$\W(n)$ can be defined as an universal algebra with $2n$ generators $\{x^1\ldots x^n\}\cup\{P_1\ldots P_n\}$ satisfying
the following set of commutation relations
\begin{gather}  \label{Weyl}
P_{\mu }x^{\nu}-x^\nu P_\mu = \delta _{\mu}^\nu\, 1, \qquad  x^{\mu }x^{\nu } - x^{\nu }x^{\mu }
  = P_{\mu }P_{\nu } - P_{\nu }P_{\mu } =0\ .
\end{gather}
for $\mu, \nu =1\ldots n$.

It is worth to underline that the Weyl-Heisenberg algebra as defined above is not an enveloping algebra of some Lie algebra.  More precisely, in contrast to the  Lie algebra case, Weyl-Heisenberg algebra have no finite dimensional (i.e matrix) representations. One can check it by taking the trace of the basic commutation relation $[x,p]=1$ which leads to the contradiction.
Much in the same way one can set
\begin{proposition}
There is no bialgebra structure which is compatible with the commutation relations (\ref{Weyl}).
\end{proposition}
The proof is trivial: applying the counit $\epsilon$ to both sides of the first commutator in (\ref{Weyl}) leads to a contradiction since $\epsilon (1)=1$.

The best known representations are given on the space of (smooth) functions on $\mathds{R}^n$ in terms of multiplication and differentiation operators, i.e. $P_\mu= {\partial\over \partial x^\mu}$. For this reason one can identify Weyl-Heisenberg algebra with an algebra of linear differential operators on $\mathds{R}^n$ with polynomial coefficients. In physics, after taking a suitable real structure, it is known as an algebra of the canonical commutation relations. Hilbert space representations of these algebras play a central role in Quantum Mechanics while their counterpart with infinitely many generators (second quantization) is a basic tool in Quantum Field Theory.

 A possible deformation of Weyl-Heisenberg algebras have been under investigation \cite{Pillin}, and it turns out that there is no non-trivial deformations of the above algebra within a category of algebras. However the so-called q-deformations have been widely investigated, see e.g. \cite{Pillin,Wess,Lavagno}.

Another obstacle is that the standard, in the case of Lie algebras, candidate for undeformed (primitive) coproduct
\be\label{primitive}
\Delta_0(a)=a\otimes 1+ 1\otimes a
\ee
$a\in \{x^1\ldots x^n\}\cup\{P_1\ldots P_n\}$ is also incompatible with (\ref{Weyl}).
It makes additionally impossible to determine a bialgebra structure on the Weyl-Heisenberg algebras.

However one could weaken the notion of bialgebra and consider unital non-counital bialgebras equipped with 'half-primitive' coproducts \footnote{These formulae were announced to us  by S. Meljanac and D. Kovacevic in the context of Weyl-Heisenberg algebra.}, left or right:
\be \label{half_prim}
\Delta^L_0(x)=x\otimes 1;\qquad\Delta^R_0(x)=1\otimes x
\ee
on $\W(n)$. In contrast to (\ref{primitive}) which is valid only on generators, the formulae (\ref{half_prim}) preserve their form for all elements of the algebra.

Moreover, such coproducts turn out to be applicable also to larger class of deformed  coordinate algebras  (quantum spaces \cite{DFR1},\cite{DFR2}) being, in general, defined by commutation relations of the form
\be\label{q-space}
x^\mu x^\nu - x^\mu x^\nu = \theta^{\mu\nu}+\theta^{\mu\nu}_\lambda x^\lambda+\theta^{\mu\nu}_{\rho\sigma}x^\rho x^\sigma +\ldots
\ee
for constant parameters $\theta^{\mu\nu}, \theta^{\mu\nu}_\lambda,\theta^{\mu\nu}_{\lambda\rho},\ldots$ . Of course, one has to assume that the number of components on the right hand side of (\ref{q-space}) is finite.
\begin{proposition}
The left (right)-primitive coproduct determines a non-counital bialgebra structure on an arbitrary associative unital algebra. In particular, one can consider a class of algebras defined by the commutation relations
(\ref{q-space}).
\end{proposition}
\begin{remark}
Such deformed algebra provides a deformation quantization of $\mathds{R}^n$ equipped
with the Poisson structure:
\be
\{x^\mu,x^\nu\}=\theta^{\mu\nu}(x)=\theta^{\mu\nu}+\theta^{\mu\nu}_\lambda x^\lambda+\theta^{\mu\nu}_{\rho\sigma}x^\rho x^\sigma +\ldots
\ee
represented by Poisson bivector $\Theta=\theta^{\mu\nu}(x)\partial_\mu\wedge \partial_\nu$.\end{remark}
Particularly, one can get the so-called theta-deformation:
\be\label{theta}
[x^\mu,x^\nu]=\theta^{\mu\nu}
\ee
which can be obtained via  twisted deformation by means of Poincar\'{e} Abelian twist:
$$\F=exp(\theta^{\mu\nu}P_\mu\wedge P_\nu)$$
The same twist provides also $\theta-$ deformed Poincar\'{e} Hopf algebra as a symmetry group, i.e. the quantum group with respect to which (\ref{theta}) becomes a covariant quantum space \footnote{
Note that the twist deformation requires h-adic extension.} .

Another way to omit counital coalgebra problem for (\ref{Weyl}) relies on introducing the
central element $C$ and replacing the commutation relations (\ref{Weyl}) by the following Lie algebraic ones
\begin{gather}  \label{Heisenberg}
\left[ P_{\mu },x^{\nu }\right]=-\imath \delta _{\mu}^\nu C, \qquad \left[
x^{\mu },x^{\nu }\right]=\left[C ,x^{\nu }\right]=\left[ P_{\mu },P_{\nu }%
\right]=\left[C ,P_{\nu }\right]=0.
\end{gather}
The relations above determine $(2n+1)$-dimensional Lie algebra  of
rank $n+1$ which we shall call Heisenberg-Lie algebra $\mathfrak{hl}(n)$. This algebra can be described as a central extension of the Abelian Lie algebra
$\mathfrak{ab}(x^1,\ldots , x^n, P_1, \ldots, P_n)$. Thus Heisenberg algebra
can be now def\/ined as an enveloping algebra $\U_{\mathfrak{hl}(n)}$ for (\ref{Heisenberg}). There is no problem to introduce Hopf algebra structure with the primitive coproduct (\ref{primitive}) on the generators $\{x^1,\ldots , x^n, P_1, \ldots, P_n, C\}$.
This type of extension  provides a
starting point for Hopf algebraic deformations, e.g. quantum group framework is considered in~\cite{LukMinn}, \cite{Bonechi}, standard and nonstandard deformations are presented e.g. in \cite{tw_Heis} while deformation quantization formalism is developed in \cite{Sorace}. As a trivial example of quantum deformations of the Lie algebra (\ref{Heisenberg}) one can consider the maximal Abelian twist of the form:
\begin{equation}
\F=exp(ih\theta^{\mu\nu}P_\mu\wedge P_\nu)exp(\lambda^\mu P_\mu\wedge C)
\end{equation}
$\theta^{\mu\nu}, \lambda^\mu$-are constants (parameters of deformation).
It seems to us, however, that there are no enough strong physical motivations for studying  deformation problem for such algebras. Therefore we shall focus on possibilities of  relaxing some algebraic conditions in the definition of bicrossproduct bialgebra in order to obey the case of Weyl-Heisenberg algebra as it is defined by (\ref{Weyl}).\\

\section{Crossed product and coproduct}

\textbf{Crossed product algebras}\\
Let $\mathcal{H}= \mathcal{H}\left( m_{\mathcal{H}},\Delta _{\mathcal{H}},\epsilon_{\mathcal{H}}, 1_{\mathcal{H}}\right)$  be a (unital and counital) bialgebra\footnote{It means that at the moment we are not interested in the full Hopf algebra structure including antipodes $ S_\H$.} and $\mathcal{A}=\mathcal{A}\left( m_{\mathcal{A}}, 1_{\mathcal{A}}\right)$ be  an (unital)
algebra.
\begin{definition}
\label{module} A (left) $\mathcal{H}$-module algebra $\mathcal{A}$ over a
Hopf algebra $\mathcal{H}$ is an algebra $\mathcal{A}$ which is a left $%
\mathcal{H}$-module such that $m_\A:\mathcal{A}\otimes\mathcal{A}%
\shortrightarrow\mathcal{A}$ and $1_\A:\mathbb{C}\shortrightarrow\mathcal{A}$
are left $\mathcal{H}$-module homomorphisms. If $\triangleright:\mathcal{H}%
\otimes\mathcal{A}\rightarrow\mathcal{A}$ denotes (left) module action $%
L\triangleright f$ of $L\in\mathcal{H}$ on $f\in\mathcal{A}$ the following
compatibility condition is satisfied: 
\begin{gather}  \label{genLeibniz}
L\triangleright (f\cdot g)=(L_{(1)}\triangleright f)\cdot
(L_{(2)}\triangleright g)
\end{gather}
for $L\in\mathcal{H}$, $f,g\in\mathcal{A}$ and $L\triangleright
1=\epsilon(L)1$, $1\triangleright f=f$ (see, e.g., \cite{Klimyk, Kassel}).%
\newline
And analogously for right $\mathcal{H}$-module algebra $\mathcal{A}$ the
condition:
\begin{equation*}
(f\cdot g)\triangleleft L=(f\triangleleft L_{(1)})\cdot (g\triangleleft
L_{(2)})
\end{equation*}
is satisfied, with (right)-module action $\triangleleft :\mathcal{A}\otimes
\mathcal{H}\rightarrow \mathcal{A}$; for $L\in \mathcal{H}$, $f,g\in
\mathcal{A}$, $1\triangleleft L=\epsilon (L)1$, $f\triangleleft 1=f$.
\end{definition}

\begin{definition} Let $\mathcal{A}$ be a left $\mathcal{H}$-module algebra.
Crossed product algebra $\A\rtimes\H$ is an algebra determined on the vector space $\A\otimes\H$ by the multiplication:
\begin{gather}\label{smash}
(f\otimes L)\rtimes(g\otimes M)=f(L_{(1)}\triangleright g)\otimes L_{(2)}M
\end{gather}
\end{definition}
Obviously, it contains algebras $\A\ni a\rightarrow a\otimes 1$ and $\H\ni L\rightarrow 1\otimes L$ as subalgebras.
Similarly, in the case of right $\mathcal{H}$-module algebra $\mathcal{A}$ the crossed product $\mathcal{%
H\ltimes A}$ is determined on the vector space $\mathcal{H}\otimes\mathcal{A}$ by:
$(L\otimes f)\ltimes(M\otimes g)=LM_{(1)}\otimes (f\triangleleft M_{(2)})g$. The trivial action $M\triangleright f= \epsilon(L)f$ reconstructs the ordinary tensor product of two algebras $\A\otimes\H$ with trivial cross-commutation relations $[f\otimes 1, 1\otimes M]=0$.

As an example we take  Weyl-Heisenberg algebra introduced above (\ref{Weyl}). For this purpose one considers two copies of Abelian $n-$dimensional Lie algebras: $\mathfrak{ab}(P_1,\ldots,P_{n})$, $\mathfrak{ab}(x^1,\ldots, x^{n})$ together with  the corresponding universal enveloping  algebras $\U_{\mathfrak{ab}(P_1,\ldots,P_{n})}$  and $\U_{\mathfrak{ab}(x^1,\ldots,x^{n})}$. Alternatively both algebras are isomorphic to the universal commutative algebras with $n$ generators (polynomial algebras). These two algebras  constitute a dual pair of Hopf algebras. Making use of primitive coproduct on generators of $\U_{\mathfrak{ab}(P_1,\ldots,P_{n})}$  we extend the (right) action implemented by duality map
\begin{gather}  \label{actionP2}
x^\nu\triangleleft P_{\mu}=\delta_\mu^\nu, \qquad 1\triangleleft P_{\mu}=0
\end{gather}
to the entire algebra  $\U_{\mathfrak{ab}(x^1,\ldots,x^{n})}$. Thus
 $W(n)=\U_{\mathfrak{ab}(P_1,\ldots,P_{n})}\ltimes\U_{\mathfrak{ab}(x^1,\ldots,x^{n})}$.

Similarly, the Heisenberg-Lie algebra can be obtained in the same way provided slight modifications in the action:
\begin{equation}\label{actionP3}
x^\nu\triangleleft P_{\mu}=\delta_\mu^\nu C, \qquad C\triangleleft
P_{\mu}=0
\end{equation}
It gives  $\U_{\mathfrak{hl}(n)}=\U_{\mathfrak{ab}(P_1,\ldots,P_{n})}\ltimes\U_{\mathfrak{ab}(x^1,\ldots,x^{n},C)}$.

\textbf{Crossed coproduct coalgebras} \cite{MR_bicross2},\cite{Klimyk}\newline
The dual concept to the action of an algebra (introduced in def. \ref{module}%
) is the \textit{coaction} of a coalgebra. Let now $\mathcal{A}=\mathcal{A}\left( m_{\mathcal{A}}, \Delta _{\mathcal{A}},\epsilon_{\mathcal{A}}, 1_{\mathcal{A}}\right)$ be a bialgebra and $\mathcal{H}= \mathcal{H}\left( \Delta_{\mathcal{H}}, \epsilon_{\mathcal{H}}\right)$  be a coalgebra. The left coaction of the bialgebra $
\mathcal{A}$ over the coalgebra $\mathcal{H}$ is defined as linear map: $\beta :%
\mathcal{H}\rightarrow \mathcal{A}\otimes \mathcal{H};$ with the following Sweedler type
notation: $\beta \left( L\right) = L^{\left( -1\right) }\otimes
L^{\left( 0\right) }$, where $L^{\left( -1\right) }\in\mathcal{A}$ and $%
L^{\left( 0\right) }\in\mathcal{H}$, $\beta(1_\H)=1_\A\otimes 1_\H$.

\begin{definition}
\label{beta} We say that $\mathcal{H}$ is left $\mathcal{A}$ -\textbf{%
comodule coalgebra} with the structure map $\beta :\mathcal{H}\rightarrow
\mathcal{A}\otimes \mathcal{H}$ if this map satisfies the following two
conditions: $\forall f,g\in \mathcal{A};L,M\in \mathcal{H}$

1)
\begin{equation}  \label{com_1}
\left( id_{\mathcal{A}}\otimes \beta \right)\circ \beta =\left( \Delta_{%
\mathcal{A}} \otimes id_{\mathcal{H}}\right)\circ \beta
\end{equation}
which can be written as: $L^{\left( -1\right) }\otimes (L^{\left( 0\right)}
)^{\left( -1\right) }\otimes (L^{\left( 0\right)} )^{\left( 0\right)
}=\left( L^{\left( -1\right) }\right) _{\left( 1\right) }\otimes \left(
L^{\left( -1\right) }\right) _{\left( 2\right) }\otimes L^{\left( 0\right) }$%
\newline
and $(\epsilon_{\mathcal{A}}\otimes id_{\mathcal{H}})\circ\beta=id_{\mathcal{%
H}}$ which reads as: $\epsilon _{\mathcal{A}}\left( L^{\left( -1\right)
}\right) L^{\left( 0\right) }= L;$

2) Additionally it satisfies comodule coaction structure (comodule coalgebra
conditions):
\begin{equation}  \label{com_2}
L^{\left( -1\right) } \epsilon _{\mathcal{H}}\left( L^{\left( 0\right)
}\right) =1_{\mathcal{A}}\epsilon _{\mathcal{H}}\left( L\right)
\end{equation}
\begin{equation}
L^{\left( -1\right) }\otimes \left( L^{\left( 0\right) }\right) _{\left(
1\right) }\otimes \left( L^{\left( 0\right) }\right) _{\left( 2\right)
}=\left( L_{\left( 1\right) }\right) ^{\left( -1\right) }\left( L_{\left(
2\right) }\right) ^{\left( -1\right) }\otimes \left( L_{\left( 1\right)
}\right) ^{\left( 0\right) }\otimes \left( L_{\left( 2\right) }\right)
^{\left( 0\right) }
\end{equation}
\end{definition}

Left $\mathcal{A}$-comodule coalgebra is a bialgebra $\mathcal{H}$ which is
left $\mathcal{A}$-comodule such that $\Delta _{\mathcal{H}}$ and $\epsilon
_{\mathcal{H}}$ are comodule maps from definition \ref{beta}. 

For such a left $\mathcal{A}$ - comodule coalgebra $\mathcal{H}$, the vector
space $\mathcal{H}\otimes \mathcal{A}$ becomes a (counital) coalgebra with the
comultiplication and counit defined by:
\begin{equation}  \label{copcom}
\Delta_\beta \left( L\otimes f\right) =\sum L_{\left( 1\right) }\otimes \left(
L_{\left( 2\right) }\right) ^{\left( -1\right) }f_{\left( 1\right) }\otimes
\left( L_{\left( 2\right) }\right) ^{\left( 0\right) }\otimes f_{\left(
2\right) }
\end{equation}
\begin{equation}  \label{epsiloncom}
\epsilon \left( L\otimes f\right) =\epsilon _{\mathcal{H}}\left( L\right)
\epsilon _{\mathcal{A}}\left( f\right)
\end{equation}
$L\in \mathcal{H};f\in \mathcal{A}.$

This coalgebra is called the \textbf{left crossed product coalgebra} and it
is denoted by $\mathcal{H}\rtimes ^{\beta }\mathcal{A}$ or $\mathcal{H}%
\rtimes \mathcal{A}.$ One should notice that:
\begin{equation*}
\Delta_\beta(L\otimes 1_{\mathcal{A}})=\left(L_{(1)}\otimes
(L_{(2)})^{(-1)}\right)\otimes\left((L_{(2)})^{(0)}\otimes 1_{\mathcal{A}%
}\right)=L_{(1)}\otimes \beta(L_{(2)})\otimes 1_{\mathcal{A}}
\end{equation*}
and
\begin{equation*}
\Delta_\beta(1_{\mathcal{H}}\otimes g)=\left(1_{\mathcal{H}}\otimes
g_{(1)}\right)\otimes\left(1_{\mathcal{H}}\otimes g_{(2)}\right)
\end{equation*}
i.e. $\Delta_\beta(\tilde f) =\tilde f_{(1)}\otimes\tilde f_{(2)}$, where $\tilde f= 1_\H\otimes f$. Moreover for the trivial choice
\be\label{trivial}
\beta_{trivial} (M)=1_\A\otimes M
\ee
one also gets
\be
\Delta_\beta(\tilde M)=\tilde M_{(1)}\otimes\tilde M_{(2)}
\ee
where $\tilde M=M\otimes 1_\A$. This implies that both coalgebras are subcoalgebras in $\mathcal{H}\rtimes \mathcal{A}.$
\begin{remark}\label{r7}
Let us assume for a moment that the coalgebra $\H$ has no counit. Leaving remaining assumptions in the same form and skipping ones containing $\epsilon_\H$ we can conclude that the resulting coalgebra  $\mathcal{H}\rtimes ^{\beta }\mathcal{A}$ has no counit (\ref{epsiloncom}) as well. In other words all other elements of the construction work perfectly well.
\end{remark}

\section{Bicrossproduct construction}
Through this section let both $\mathcal{H}$ and $\mathcal{A}$ be
bialgebras. The structure of an action is useful for crossed product algebra construction and a coaction map allows us to consider crossed coalgebras. However considering both of them simultaneously we are able to perform the so-called bicrossproduct construction. %
\begin{theorem} \label{becross}
(S. Majid \cite{MR_bicross2}, Theorem 6.2.3) Let $\mathcal{H}$ and $\mathcal{A}$ be bialgebras and $%
\mathcal{A}$ is right $\mathcal{H}$-module with the structure map $\triangleleft
:\mathcal{A}\otimes \mathcal{H}\rightarrow \mathcal{A}$. And $\mathcal{H}$
is left $\mathcal{A}$-comodule coalgebra with the structure map\newline
$\beta :\mathcal{H}\rightarrow \mathcal{A}\otimes \mathcal{H}$, $\beta
\left( L\right) =L^{\left( -1\right) }\otimes L^{\left( 0\right) }$ (cf.
def. \ref{beta}).\\
Assume further the following compatibility conditions:\newline
(A)
\begin{equation}  \label{bicrossA1}
\Delta_{\mathcal{A}} \left( f\triangleleft L\right) =\sum \left(
f\triangleleft L\right) _{\left( 1\right) }\otimes \left( f\triangleleft
L\right) _{\left( 2\right) }= \left( f_{\left( 1\right) }\triangleleft
L_{\left( 1\right) }\right) \left( L_{\left( 2\right) }\right) ^{\left(
-1\right) }\otimes f_{\left( 2\right) }\triangleleft \left( L_{\left(
2\right) }\right) ^{\left( 0\right) }
\end{equation}
\begin{equation}  \label{bicrossA2}
\epsilon _{\mathcal{A}}\left( f\triangleleft L\right) =\epsilon _{\mathcal{A}%
}\left( f\right) \epsilon _{\mathcal{H}}\left( L\right)
\end{equation}
(B)
\begin{equation}  \label{bicrossB}
\beta \left( LM\right) =\left( LM\right) ^{\left( -1\right) }\otimes \left(
LM\right) ^{\left( 0\right) }=\sum \left( L^{\left( -1\right) }\triangleleft
M_{\left( 1\right) }\right) \left( M_{\left( 2\right) }\right) ^{\left(
-1\right) }\otimes L^{\left( 0\right) }\left( M_{\left( 2\right) }\right)
^{\left( 0\right) }
\end{equation}
\begin{equation}  \label{bicrossB2}
\beta(1_{\mathcal{H}})\equiv\left( 1_{\mathcal{H}}\right) ^{\left( -1\right) }\otimes \left( 1_{\mathcal{%
H}}\right) ^{\left( 0\right) }=1_{\mathcal{A}}\otimes 1_{\mathcal{H}}
\end{equation}
(C)
\begin{equation}  \label{bicrossC}
\left( L_{\left( 1\right) }\right) ^{\left( -1\right) }\left( f\triangleleft
L_{\left( 2\right) }\right) \otimes \left( L_{\left( 1\right) }\right)
^{\left( 0\right) }=\left( f\triangleleft L_{\left( 1\right) }\right) \left(
L_{\left( 2\right) }\right) ^{\left( -1\right) }\otimes \left( L_{\left(
2\right) }\right) ^{\left( 0\right) }
\end{equation}
hold. Then the crossed product algebra $\mathcal{H}\ltimes\mathcal{A}$,
i.e. tensor algebra $\mathcal{H}\otimes\mathcal{A}$ equipped with algebraic:
\begin{equation*}
(L\otimes f)\cdot(M\otimes g)=LM_{(1)}\otimes (f\triangleleft
M_{(2)}g)\qquad\qquad\qquad(product)
\end{equation*}
\begin{equation*}
1_{\mathcal{H}\ltimes\mathcal{A}}=1_\mathcal{H}\otimes 1_\mathcal{A}
\qquad\qquad\qquad\qquad\qquad\qquad\qquad(unity)
\end{equation*} and coalgebraic
\begin{equation}\label{cop_bi}
\Delta_\beta(L\otimes f)=\left(L_{(1)}\otimes (L_{(2)})^{(-1)}
f_{(1)}\right)\otimes\left((L_{(2)})^{(0)}\otimes
f_{(2)}\right)\qquad(coproduct)
\end{equation}
\begin{equation*}
\epsilon(L\otimes
f)=\epsilon_\H(L)\epsilon_\A(f)\qquad\qquad\qquad\qquad\qquad\qquad\qquad(counit)
\end{equation*}
sectors becomes a bialgebra. Following \cite{MR_bicross1,MR_bicross2} one calls it bicrossproduct bialgebra and denotes as $\mathcal{H}%
\Join\mathcal{A}$. Moreover if the initial algebras are Hopf algebras then
introducing the antipode:
\begin{equation*}
S(L\otimes f)=(1_\mathcal{H}\otimes S_\A(L^{(-1)}f))\cdot(S_\H(L^{(0)})\otimes 1_%
\mathcal{A})\qquad\qquad(antipode)
\end{equation*}
it becomes bicrossproduct Hopf algebra $\mathcal{H}\Join\mathcal{A}$ as well.
\end{theorem}
\begin{example}
Primitive Hopf algebra structure on $\U_{\mathfrak{hl}(n)}$ can be obtained via bicrossproduct construction.
Take $\U_{\mathfrak{ab}(P_1,\ldots,P_{n})}$ as left $\U_{\mathfrak{ab}(x^1,\ldots,x^{n},C)}$ comodule algebra with the trivial coaction map: $\beta(P_{\mu})=1\otimes P_{\mu}$. Taking into account the action (\ref{actionP3})
all assumptions from the previous theorem are fulfilled.
Thus due  to the formula (\ref{cop_bi})
one obtains the following coalgebraic structure:\\ $\Delta \left( \tilde{P}_{\nu
}\right) =\tilde{P}_{\nu }\otimes 1+1\otimes \tilde{P}_{\nu };\qquad\Delta
\left( \tilde{x}^\nu\right) =\tilde{x}^\nu\otimes
1+1\otimes \tilde{x}^\nu;\qquad\Delta(\tilde{C})=\tilde{C}\otimes 1+1\otimes \tilde{C}$\newline
with canonical Hopf algebra embeddings: $1\otimes P_{\nu }\rightarrow \tilde{P}_{\nu
};\ {x}^\nu\otimes 1\rightarrow \tilde{x}^\nu;\ C\otimes 1\rightarrow \tilde{C}$.
\end{example}

The last example suggests the following more general statement:
\begin{proposition}
Let $\U_{\mathfrak{g}}$ and $\U_{\mathfrak{h}}$ be two enveloping algebras corresponding to two finite dimensional Lie algebras $\mathfrak{g},\mathfrak{h}$, both equipped in the primitive coalgebra structure (i.e. the coproduct $\Delta(x)=x\otimes 1+1\otimes x$ for $x\in {\mathfrak{g}}\cup {\mathfrak{h}}$). Assume that the (right) action of $\U_{\mathfrak{g}}$ on $\U_{\mathfrak{h}}$ is of Lie type, i.e. it is implemented by Lie algebra action: $h_a\triangleleft g_i =c_{ia}^b h_b$ in some basis $g_i$, $h_a$, where $c_{ia}^b$ are numerical constants.
Then one can always define the primitive Hopf algebra structure on $\U_{\mathfrak{g}}\ltimes\U_{\mathfrak{h}}$ by using 
bicrossproduct construction with the trivial co-action map: $\beta_{trivial}(g_i)=1\otimes g_i $. \end{proposition}
However from our point of view the most interesting case is deformed one.
To this aim let us remind  bicrossproduct construction for $\kappa$-Poincar\'{e} quantum group. In contrast to the original construction presented in \cite{MR} the resulting Hopf algebra structure will be determined in the classical Poincar\'{e} basis.
\begin{example}
We take as the first component enveloping algebra of 4-dimensional Lorentz Lie algebra $\mathfrak{o}\left( 1,3\right)$, closed in  h-adic topology, i.e. $\mathcal{H}=\mathcal{U}_{\mathfrak{o}\left( 1,3\right)}[[h]]$ with the primitive (undeformed) coalgebra structure (\ref{primitive}).
As the second component we assume Hopf algebra of translations $\mathcal{A}=\U_{\mathfrak{ab}(P_1,P_2,P_3,P_4)}[[h]] $ with nontrivial coalgebraic sector:
\begin{equation}
\Delta _{\kappa }\left( P_{i}\right) =P_{i}\otimes \left(hP_{4}+\sqrt{%
1-h^{2} P^{2}}\right)+1\otimes P_{i} \ ,\qquad i=1,2,3
\end{equation}
\begin{equation}
\Delta _{\kappa }\left( P_{4}\right) =P_{4}\otimes \left(h P_{4}+\sqrt{%
1-h^{2} P^{2}}\right)+\left(hP_{4}+\sqrt{1-h^{2} P^{2}}\right)^{-1}\otimes
P_{4} +hP _{m}\left(hP_{4}+\sqrt{1-h^{2} P^{2}}\right)^{-1}\otimes P^{m},
\label{copP0}
\end{equation}
here $P^2=P_\mu P^\mu$ and $\mu=1,\ldots ,4$. Observe that one deals here with formal power series in the formal parameter $h$ (cf. \cite{BP_sigma}).
Now $\U_{\mathfrak{ab}(P_1,\ldots,P_4)}[[h]] $ is a right $\mathcal{U}_{\mathfrak{o}\left( 1,3\right)}[[h]]$ module algebra implemented by the  classical (right) action:
\begin{gather}  \label{class_right_action}
 P_{k}\triangleleft M_{j}=\imath \epsilon
_{jkl}P_{l},\qquad P_{4}\triangleleft M_{j}=0, \\
P_{k}\triangleleft N_{j} =- \imath \delta _{jk}P_{4},\qquad
P_{4}\triangleleft N_{j}=-\imath P_{j}  \label{class_right_action2}
\end{gather}
Conversely, $\mathcal{U}_{\mathfrak{o}\left( 1,3\right) }[[h]]$ is a left $\U_{\mathfrak{ab}(P_1,\ldots,P_4)}[[h]] $ - comodule coalgebra with (non-trivial) structure map defined on generators as follows:  
\begin{equation}\label{beta_kap1}
\beta_{\kappa} \left( M_{i}\right) =1\otimes M_{i}
\end{equation}
\begin{equation}\label{beta_kap2}
\beta_{\kappa} \left( N_{i}\right) =\left(h P_{4}+\sqrt{1-h^{2} P^{2}}%
\right)^{-1}\!\otimes N_{i}-h\epsilon _{ijm} P_{j}\left(h P_{4}+\sqrt{%
1-h^{2} P^{2}}\right)^{-1}\!\otimes M_{m}
\end{equation}
and then extended to the whole universal enveloping algebra. Such choice guarantees that all the conditions (\ref{bicrossA1}-\ref{bicrossC}) are fulfilled.
Thus  the structure obtained via bicrossproduct construction  constitutes  Hopf algebra $\mathcal{U}%
_{\mathfrak{o}\left( 1,3\right) }[[h]]\Join\U_{\mathfrak{ab}(P_1,\ldots,P_4)}[[h]] $ which has classical algebraic sector
while coalgebraic one reads as introduced in \cite{BP2,BP_sigma}.
\end{example}

\begin{remark} We are in position now to extend remark (\ref{r7}) to the bicrossproduct case. Again we have to neglect counit
on the bialgebra $\H$. As a result one obtains unital and non-counital bialgebra $\mathcal{H}\Join\mathcal{A}$.
\end{remark}
As an illustrative example of such  constrution one can consider Weyl-Heisenberg algebra (\ref{Weyl}).
The algebra of translations $\U_{\mathfrak{ab}(P_1,\ldots,P_n)}$ is taken  with primitive coproduct.  Non-counital bialgebra of spacetime (commuting) coordinates $\U_{\mathfrak{ab}(x^1,\ldots,x^{n})}$ is assumed to posses half-primitive coproduct. The action is the same as in (\ref{actionP2}) while coaction is assumed to be trivial. As a final result one gets non-counital and non-cocommutative bialgebra structure on $\W(n)$:  $\Delta \left( \tilde{P}_{\nu
}\right) =\tilde{P}_{\nu }\otimes 1+1\otimes \tilde{P}_{\nu };\qquad\Delta
\left( \tilde{x}^\nu\right) =\tilde{x}^\nu\otimes 1$ , where $1\otimes P_{\nu }\rightarrow \tilde{P}_{\nu
};\ {x}^\nu\otimes 1\rightarrow \tilde{x}^\nu$.
\section{Conclusions}
It is still an open problem what kind of deformations can be encoded in the bicrossproduct construction.
For example, in the class of twisted deformation we were unable to find a single case obtained by means of such construction.
Nevertheless $\kappa$-deformation of the Poincar\'{e} Lie algebra is one of few  examples of quantization  for which bicrossproduct description works perfectly. More sophisticated examples can be found in \cite{more1}-\cite{more3}.
Moreover, it has been proved in \cite{BP_sigma} that large class of deformations of the Weyl-Heisenberg algebra $\W(n)$ can be obtained as a (non-linear) change of generators in its h-adic extension $W(n)[[h]]$. Therefore our results concerning construction of non-counital bialgebra structure extend automatically to these cases.

\section*{Acknowledgment}
The authors acknowledge discussions with  D. Kovacevic and S. Meljanac. The work of A.P. was supported by the Polish Ministry of
Science and  Higher Education grant NN202 238540. AB is supported by the Polish National Science Center project 2011/01/B/ST2/03354.


\end{document}